\documentclass{article}

\usepackage{amssymb,amsfonts,amsmath,stmaryrd}
\usepackage{cite,enumerate,float,indentfirst}
\usepackage{color}
\usepackage{tikz}
\usetikzlibrary{arrows,snakes,backgrounds}
\usepackage{hyperref}

\def\be{\begin{eqnarray}}
\def\ee{\end{eqnarray}}
\def\nn{\nonumber}

\def\p{\partial}
\def\tr{{\rm tr}\,}

\def\SP{{\rm SP}\,}
\def\OP{{\rm OP}\,}

\definecolor{red}{rgb}{1,0,0}
\definecolor{orange}{rgb}{1,0.5,0}
\definecolor{violet}{rgb}{0.7,0,1}

\hoffset= - 1.0in         

\begin{document}

\begin{center}
\begin{small}
\hfill MIPT/TH-18/21\\
\hfill FIAN/TD-15/21\\
\hfill IITP/TH-22/21\\
\hfill ITEP/TH-32/21\\

\end{small}
\end{center}

\vspace{.5cm}

\begin{center}
\begin{Large}\fontfamily{cmss}
\fontsize{15pt}{27pt}
\selectfont
	\textbf{Spin Hurwitz theory and Miwa transform for the Schur Q-functions  }
	\end{Large}
	
\bigskip \bigskip

\begin{large}
A. Mironov$^{b,c,d}$\footnote{mironov@lpi.ru; mironov@itep.ru},
A. Morozov$^{a,c,d}$\footnote{morozov@itep.ru},
A. Zhabin$^{a,d}$\footnote{alexander.zhabin@yandex.ru}
\end{large}

\bigskip

\begin{small}
$^a$ {\it MIPT, Dolgoprudny, 141701, Russia}\\
$^b$ {\it Lebedev Physics Institute, Moscow 119991, Russia}\\
$^c$ {\it Institute for Information Transmission Problems, Moscow 127994, Russia}\\
$^d$ {\it ITEP, Moscow 117218, Russia}
\end{small}
 \end{center}

\bigskip

\begin{abstract}
Schur functions are the common eigenfunctions of generalized
cut-and-join operators
which form a closed algebra.
They can be expressed as differential operators in time-variables
and also through the eigenvalues of auxiliary $N\times N$ matrices $X$,
known as  Miwa variables.
Relevant for the cubic Kontsevich model and also for spin Hurwitz theory
is an alternative set of Schur Q-functions.
They appear in representation theory of the Sergeev group, which is a substitute of the symmetric group,
related to the queer Lie superalgebras $\mathfrak{q}(N)$.
The corresponding spin $\hat{\cal W}$-operators were recently found in terms of time-derivatives,
but a substitute of the Miwa parametrization remained unknown, which
is an essential complication for the matrix model technique and further developments.
We demonstrate that the Miwa representation, in this case, involves a fermionic matrix $\Psi$
in addition to $X$, but its realization using supermatrices is {\it not} quite naive.
\end{abstract}

\section{Introduction}

The basic question of quantum field theory is to find the correlators,
the values of path integrals with a given action and appropriate
basis of probe functions.
In the simplest case of Gaussian matrix models, this basis is provided \cite{MM}
by the Schur functions \cite{Mac},
and this has far going implications,
in  particular,  for the theory of Hurwitz numbers \cite{MMN1}.
Nowadays, it is known \cite{MMKon,MMNspin} that the two direct generalizations, those
to the cubic Kontsevich model \cite{K,GKM} and to the theory of the spin Hurwitz numbers \cite{EOP,G,LP,MMNspin}
are related to the averages of the somewhat different Schur Q-functions \cite{Mac}.
It is a challenging problem to extend our knowledge about
the averages of Schur functions to those of the Schur Q-functions.
In this paper, we address one of the subjects of this rich and important story.

We discuss one of the basic objects related to the appropriate basis:
a set of commuting differential operators that have the elements of the basis as their eigenfunctions.
In the case of the Schur functions (hence, of Gaussian matrix models), these operators are the generalized cut-and-join operators $\hat W_\Delta$ enumerated by the Young diagrams $\Delta$ \cite{GJ,MMN1,MMN2}.
They are made from the Casimir operators of the linear algebra $gl(\infty)$,
and have  as their common eigenfunctions all the characters of the linear group $GL(N)$ written in the $N$-independent form, i.e.
the Schur functions written as functions of power sums.
These operators  form a linear basis in the center of the universal enveloping algebra $U_{gl(\infty)}$.
A useful representation of  $\hat W_\Delta$ is through the matrix differential operators \cite{MMN1}
made from the Miwa matrix-variable $X$ and realizing the regular representation \cite{RR,GKMMMO} of $GL(N)$ at particular $N$.
They form an algebra isomorphic to the Ivanov-Kerov algebra \cite{IK}.

In this letter, we extend this construction to the Schur Q-functions, and construct the generalized cut-and-join operators $\hat {\cal W}_\Delta$ that have the Schur Q-functions as their eigenfunctions. They can be presented as the matrix differential operators made from the operators which realize the regular representation of the supergroup $\mathfrak{Q}(N)$ associated with the superalgebra $\mathfrak{q}(N)$ \cite{Superalg}, so that $\hat {\cal W}_\Delta$ form a linear basis in the center of the universal enveloping algebra $U_{\mathfrak{q}(N)}$, in particular, in the limit of $N\to\infty$ as above. The Schur Q-functions are characters of the supergroup $\mathfrak{Q}(N)$ and, hence, are eigenfunctions of $\hat {\cal W}_\Delta$. The crucial subtlety is a clever choice of proper Miwa variables, since $\hat {\cal W}_\Delta$ involves both {\it even} and {\it odd} variables, which we call $X$ and $\Psi$,
and one has to choose the variables in such a way that they also depend on the both types of variables.
In order to avoid possible misunderstanding, we use italic to distinguish Grassmannian variables from times with odd indices.

\bigskip

\section{Cut-and-join operators for Schur functions}

The Schur functions are eigenfunctions of the standard generalized cut-and-join operators \cite{MMN1}:
\be
\label{ord_W_Schurs}
\hat W_\Delta \cdot S_R =  {\phi_R(\Delta)}\cdot S_R
\ee
where the eigenvalues are expressed through the symmetric group characters $\psi_{R}(\Delta)$:
 \begin{equation}\label{phi_normalized_extended}
    \phi_{R}(\Delta) := \begin{cases}
    \phantom{AAAAAAAAAA} 0 &  \ \ \ \ \ \ \ \ \ \ \ \ |\Delta| > |R| \\
 &\\
 \displaystyle{\frac{(|R| - |\Delta| + r)!}{r! (|R| - |\Delta|)!} \cdot
 {\psi_{R}([\Delta, 1^{|R|-|\Delta|}])\over d_R\cdot z_{[\Delta, 1^{|R|-|\Delta|}]}}}
 & \ \ \ \ \ \ \ \ \ \ \ \ |\Delta| \le |R|
    \end{cases}
\end{equation}
In this formula, $r$ is the number of lines of unit length in $\Delta$, and we use the standard combinatorial quantities \cite{Fulton}: $d_R=S_R\{\delta_{k,1}\}$, which depends on the partition (Young diagram) $R$, and the order of automorphism of partition $z_\Delta$, which, in the parametrization of partition $\Delta = [1^{m_{1}}, 2^{m_{2}}, \dots ]$ is equal to $z_\Delta = \prod_{k \ge 1} k^{m_{k}} m_{k}!$.

The Schur functions,
\be
S_R\{p\}
= \sum_{\Delta\vdash |R|} d_R\phi_R(\Delta)\cdot  p_{\Delta}
\ee
are parameterized by the Young diagrams $R$, and are graded polynomials of the ``time-variables" $p_k$. Here $p_\Delta:=\prod_{s\ge 1}^{l_{\Delta}}p_{\Delta_s}$, and $l_{\Delta}$ is the length of partition (the number of lines in the Young diagram).

If restricted to ``Miwa domain" $p_k = \tr X^k$,
the Schur functions $S_R\{p_k = \tr X^k\}$ become polynomials of the $N\times N$ matrix $X$, or symmetric polynomials of eigenvalues of this matrix.
The operators $\hat W_\Delta$ acquire a very simple form in these matrix variables:
\be
\hat W_\Delta
 = \frac{1}{z_\Delta} : \prod_{s=1}^{l_{\Delta}}\hat C_{\Delta_s}:
\label{WviaX}
\ee
with $\Delta=\{\Delta_1\geq \Delta_2\geq\ldots\geq\Delta_{l_{\Delta}}>0\}$,
and the Casimir operators
\be
\hat C_{n} := \tr \hat D^n
\ee
are simple polynomials in the matrix-valued operators
\be\label{D}
\hat D_{i,j} = \sum_{a} X_{i,a} \frac{\p}{\p X_{j,a}}
\ee
The normal ordering $:\ldots :$ means that all the derivatives are put to the right.

The generalized cut-and-join operators $\hat{W}_{\Delta}$
form a  closed algebra with respect to the multiplication,
\be
\hat W_{\Delta_1} \hat W_{\Delta_2} = \sum_{\Delta_3} N_{\Delta_1\Delta_2}^{\Delta_3}
\hat W_{\Delta_3}
\ee
and the integer-valued
structure constants are the same as in the center of group algebra of symmetric group
\be
N_{\Delta_1\Delta_2}^{\Delta_3} =
z_{\Delta_3} \sum_{R\vdash |\Delta_3|} d_R^2\phi_R(\Delta_1)\phi_R(\Delta_2)\phi_R(\Delta_3)
\label{Cvspsi}
\ee
Indeed, from (\ref{ord_W_Schurs}) it follows that
\be\label{IK}
\phi_R(\Delta_1) \phi_R(\Delta_2) = \sum_{\Delta_3} N_{\Delta_1\Delta_2}^{\Delta_3}
\phi_R(\Delta_3)
\ee
for any $R$. Then, (\ref{Cvspsi}) immediately follows from this formula and the orthogonality of the characters,
\be\label{OPR}
\sum_{R}{\psi_R(\Delta)\psi_{R}(\Delta')\over z_\Delta}=\delta_{\Delta\Delta'}
\ee

\section{Generalization to Schur Q-functions}

The question is what generalizes these facts to the Schur Q-functions (the definition and details about the Schur Q-functions can be found in \cite{MMNspin}):

\bigskip

{\bf (i)} What are the operators $\hat{\cal W}_\Delta$ which satisfy
\be
\hat {\cal W}_\Delta \cdot Q_R =  {\Phi_R(\Delta)} \cdot Q_R
\label{efcalW}
\ee
with the Schur Q-functions being graded polynomials of odd time-variables $P_{2k+1}$ only,
\be\label{Fro}
Q_R\{P_k\} = \sum_{\Delta\in OP} \mathfrak{d}_R\Phi_R(\Delta)\cdot P_\Delta
\ee
for $R\in \SP$, $\Delta \in \OP$, and $\mathfrak{d}_R:=Q_R\{\delta_{k,1}\}/2$.
Here $\SP$ and $\OP$ are  the strict and odd partitions,
i.e. those with all the line lengths in the Young diagram respectively different or odd.
The eigenvalues
$\Phi_{R}(\Delta)$ are expressed through the characters of spin representations of Sergeev group $\chi_R(\Delta)$  \cite{Serg}:
\be\label{cPhi}
\Phi_R(\Delta)=\left\{\begin{array}{cl}
0\ \ \ \ \ \ \ \ \ \ \ \ \ \ \ \ \ \ \ \ \ \ \ \ \ \  &|\Delta|>|R|\\
&\\
\displaystyle{(|R|-|\Delta|+r)!\over r!(|R|-|\Delta|)!}\
\displaystyle{2^{-\frac{\delta(R)}{2}}\chi_{R}([\Delta,1^{|R|-|\Delta|}])\over \mathfrak{d}_Rz_{[\Delta,1^{|R|-|\Delta|}]}}\ \ \ \ \ \ \ \ \ \ \ \ \ \ \ \ \ \ \ \ \ \ \ &|\Delta|\le |R|
\end{array}\right.
\ee
In this formula, $r$ is again the number of lines of unit length in $\Delta$ and $\delta(R)$ is equal to $0$ for even $l(R)$ and 1 otherwise.

\bigskip

{\bf (ii)} If arguments of these Q-functions $Q_R\{P_k\}$ can be
restricted to Miwa domains $P_k = \tr X^k + \ldots$
where  operators  $\hat {\cal W}_{\Delta}$ acquire simple
expressions similar to (\ref{WviaX})?

\bigskip

{\bf (iii)} What is the algebra formed by these operators?

\section{Solution in terms of supermatrices}

In \cite{MMZh} we answered the first part {\bf (i)} of this question,
by suggesting {\it some} expressions for
$\hat {\cal W}_{\Delta}$ as somewhat sophisticated differential operators
in  $P$-variables.
In this letter, we answer the other parts,
namely, {\bf (ii)}
provide an expression (Miwa transform) for the $P$-variables
through the matrix $X$ and its fermionic counterpart $\Psi$,
\be\label{PXPsi}
\boxed{
P_k
= \frac{1}{2}\tr
\! \underbrace{
\left(\begin{array}{cc} X&\Psi \\ \Psi & X\end{array}\right)
\left(\begin{array}{cc} X&-\Psi \\ -\Psi & X\end{array}\right)
\left(\begin{array}{cc} X&\Psi \\ \Psi & X\end{array}\right)
\ldots \!}_{k \ {\rm times}}
= \frac{1}{2} \tr
\!\! \left( \prod_{\alpha=1}^{k} \left(X + (-1)^{\alpha}\Psi \right) + \prod_{\alpha=1}^{k} \left(X - (-1)^{\alpha}\Psi \right) \right)
}
 \ee
i.e. only even powers of $\Psi$ survive so that $P_k$ are {\it even} variables (while relevant indices $k$ contributing to (\ref{Fro}) are odd),
and {\bf (iii)} identify the half-integer-valued structure constants in
\be
\hat {\cal W}_{\Delta_1} \hat {\cal W}_{\Delta_2} = \sum_{\Delta_3} {\cal N}_{\Delta_1\Delta_2}^{\Delta_3}
\hat{\cal W}_{\Delta_3}
\ee
with those in the algebra of Sergeev characters
\be
{\cal N}_{\Delta_1\Delta_2}^{\Delta_3} =
 2^{-l_{\Delta_3}}\cdot z_{\Delta_3} \sum_{R\in SP} d_R^2\,
 \Phi_R(\Delta_1)\Phi_R(\Delta_2)\Phi_R(\Delta_3)
\ee
This formula follows from
\be\label{MMNspin}
\Phi_R(\Delta_1) \Phi_R(\Delta_2) = \sum_{\Delta_3} {\cal N}_{\Delta_1\Delta_2}^{\Delta_3}
\Phi(\Delta_3)
\ee
and from the orthogonality property of characters,
\be
\sum_{R\in \hbox{\footnotesize SP}}{\chi_R(\Delta)\chi_{R}(\Delta')\over 2^{\delta(R)} 2^{l_{\Delta}}
\cdot z_\Delta}=\delta_{\Delta\Delta'}
\ee

In the Miwa variables, the operators $\hat {\cal W}_\Delta(X,\Psi)$
become simple polynomials of the matrix superderivatives
\be\label{rrr}
\hat{\cal D}_{ij} \equiv \hat{\cal D}_{-i,-j} = \frac{1}{2}
\sum_{a=1}^N \left( X_{a,i}\frac{\p}{\p X_{a,j}} + \Psi_{a,i}\frac{\p}{\p \Psi_{a,j}}\right)
\nn\\
\hat{\cal D}_{i,-j} \equiv \hat{\cal D}_{-i,j} = \frac{1}{2}
\sum_{a=1}^N \left(X_{a,i}\frac{\p}{\p \Psi_{a,j}} - \Psi_{a,i}\frac{\p}{\p X_{a,j}}\right)
\ee
Define now the super-Casimir operators:
\begin{equation}
    \hat{\cal C}_{n} =
\sum_{{i_{1},i_{2}, \dots, i_{n} = -N}\atop{i_{1},i_{2}, \dots, i_{n}\ne 0}}^{N}
\hat{\cal D}_{i_{n},i_{1}} (-1)^{\bar{i}_{n}}
\ldots (-1)^{\bar{i}_{3}}
\hat{\cal D}_{i_{2},i_{3}} (-1)^{\bar{i}_{2}}
\hat{\cal D}_{i_{1},i_{2}}
\end{equation}
where $\bar{i}$ is equal to $1$ for $i>0$ and 0 otherwise.
An important property of such operators (traces) is that $\hat{\mathcal{C}}_{2n} = 0$
and non-trivial are only the odd operators $\hat{\mathcal{C}}_{2n+1}$.
Note that operator matrices in the product are transposed, in contrast with the standard case (\ref{D}). This transposition comes from the particular choice of right regular representation, (\ref{rrr}).
From these operators, we construct direct counterparts of (\ref{WviaX}):

\be
\boxed{
\hat {\cal W}_\Delta = \frac{2}{{ z}_\Delta} : \prod_{s=1}^{l_{\Delta}} \hat{\cal C}_{\Delta_{s}}:
}
\label{WviaXPsi}
\ee

\section{The check of the eigenvalues}

The $\hat{\cal W}_\Delta$ operators are polynomials of the Casimir operators of $\mathfrak{q}(n)$, hence, the Schur Q-functions are their eigenfunctions being characters of the latter with some eigenvalues $\Lambda_R(\Delta)$. Let us prove that these eigenvalues are exactly (\ref{cPhi}). To this end,
we act with the product of super-Casimir operators on the particularly simple function of $P$:
\vspace{-0.2cm}
\be\label{1}
\frac{1}{{ z}_\Delta} : \prod_{s=1}^{l_\Delta} \hat{\cal C}_{\Delta_{s}}\!:\ e^{2P_1}=
\frac{1}{{ z}_\Delta} : \prod_{s=1}^{l_\Delta} \hat{\cal C}_{\Delta_{s}}\!:\ e^{2\tr\! X}=
2^{l_\Delta} \,{P_\Delta\over z_\Delta}\,e^{2P_1}
\ee
Using the Cauchy identity for the Schur Q-functions,
\be
e^{2P_1}=\sum_{R\in \SP}2 \mathfrak{d}_R Q_R\{P_k\}
\ee
we rewrite this as
\vspace{-0.2cm}
\be\label{2}
\sum_{R\in \SP}  \mathfrak{d}_R \underbrace{\left(\frac{2}{{ z}(\Delta)}
: \prod_{s=1}^{l_\Delta} \hat{\cal C}_{\Delta_{s}}\!: \right)\cdot Q_R\{P_k\}}_{\Lambda_R(\Delta)\cdot Q_R\{P_k\}}=
2^{l_\Delta} { P_\Delta\over z_\Delta}\,e^{2P_1}
\ee
Assume, for the sake of simplicity, that the Young diagram $\Delta$ itself does not contain lines of unit length (the generalization is immediate). Then, at the r.h.s. of this formula, we get
\be
2^{l_\Delta} {P_\Delta\over z_\Delta}e^{2P_1}=\sum_m 2^{l_\Delta} {P_\Delta\over z_\Delta}{(2P_1)^m\over m!}=
\sum_m 2^{l_\Delta+m}{P_{[\Delta,1^m]}\over z_{[\Delta,1^m]}}
\ \stackrel{(\ref{iF})}{=} \
\sum_m\sum_{R\vdash |\Delta|} \mathfrak{d}_R\Phi_R([\Delta,1^m])\cdot Q_R\{P_k\}
\ \stackrel{(\ref{cPhi})}{=} \nn\\
=\sum_m\sum_{R\vdash |\Delta|+m} \mathfrak{d}_R \cdot \Phi_R(\Delta)\cdot Q_R\{P_k\}
 = \sum_{R \in \SP} \mathfrak{d}_{R} \cdot \Phi_R(\Delta)\cdot Q_R\{P_k\}
\label{3}
\ee
where we used the definition (\ref{cPhi}) and  inverse of the Fr\"obenius formula (\ref{Fro}):
\be\label{iF}
{P_\Delta\over z_\Delta}=2^{-l_\Delta}\sum_{R\vdash |\Delta|}\mathfrak{d}_R\Phi_R(\Delta)\cdot Q_R\{P_k\}
\ee
Substituting (\ref{3}) into (\ref{2}), we obtain that the eigenvalues $\Lambda_{R}(\Delta)$ of $\hat{\cal W}_\Delta$ satisfy the equation
\be
\sum_{R \in \SP} \mathfrak{d}_{R} \cdot\Big(\Lambda_R(\Delta)- \Phi_R(\Delta)\Big)\cdot Q_R\{P_k\}=0
\ee
at any $\Delta$ and $P_k$, which implies that
\be
\Lambda_R(\Delta)=\Phi_R(\Delta)
\ee

\section{Super-cut-and-join operators in terms of time-variables}

The next point is to express the operators $\hat {\cal W}_\Delta(X,\Psi)$ in the form of operators in time-variables $\hat{\cal W}_\Delta\{P\}$ that were found in \cite{MMZh}. In the latter case, one may forget about the origin of these variables (\ref{PXPsi}), and deal with them just as with formal variables.
To find the $\hat {\cal W}_\Delta\{P\}$-operators, one needs to act with (\ref{WviaXPsi}) on an arbitrary function $f(P_{1}, P_{3}, \dots)$, which depends only on odd variables $P_{2k+1}$,
and manage to express the emerging coefficients in front of the $P$-derivatives
through the $P$-variables only: naively they depend on $X$ and $\Psi$ separately.
One actually manages to do this.

The transformation from matrix derivatives to $P$-derivatives comes from the change of variables:
\begin{equation}
    \hat{\mathcal{D}}_{i,j} f(P_{1}, P_{3}, \dots) = \sum_{a=0}^{\infty} \frac{\partial f}{\partial P_{2a+1}} \cdot \hat{\mathcal{D}}_{i,j} P_{2a+1}(X,\Psi)
\end{equation}
where the action of matrix super-derivatives is
\begin{equation}\label{action_D}
    \begin{gathered}
    \hat{\mathcal{D}}_{i,j} P_{2a+1} = \frac{1}{4} (2a+1) \left( \prod_{\alpha=1}^{2a+1} \left(X + (-1)^{\alpha}\Psi \right) + \prod_{\alpha=1}^{2a+1} \left(X - (-1)^{\alpha}\Psi \right) \right)_{j,i} \\
    \hat{\mathcal{D}}_{i,-j} P_{2a+1} = \frac{1}{4} (2a+1) \left( \prod_{\alpha=1}^{2a+1} \left(X + (-1)^{\alpha}\Psi \right) - \prod_{\alpha=1}^{2a+1} \left(X - (-1)^{\alpha}\Psi \right) \right)_{j,i}
    \end{gathered}
\end{equation}
Thus, the action of $\hat{\mathcal{C}}_{1}$ is extremely simple:
\begin{equation}\label{action_C1}
    \hat{\mathcal{C}}_{1} f(P) = \left( \sum_{{i=-N}\atop{i\ne 0}}^{N} \hat{\mathcal{D}}_{i,i} \right) f(P) = \left( \sum_{i=1}^{N} 2\hat{\mathcal{D}}_{i,i} \right) f(P) = \sum_{a=0}^{\infty} (2a+1)P_{2a+1} \frac{\partial f}{\partial P_{2a+1}}
\end{equation}
Moreover, $\hat{\mathcal{C}}_{1}$ alone is enough to reproduce the $\hat{\mathcal{W}}$-operators for partitions $[1^k]$:
\begin{equation}
    \begin{gathered}
    \hat{\mathcal{W}}_{[1]} = \frac{2}{z_{[1]}} : \hat{\mathcal{C}}_{1} : = 2\hat{\mathcal{C}}_{1} \\
    \hat{\mathcal{W}}_{[1,1]} = \frac{2}{z_{[1,1]}} : \hat{\mathcal{C}}_{1}^{2} : = \hat{\mathcal{C}}_{1}^{2} - \hat{\mathcal{C}}_{1} \\
    \hat{\mathcal{W}}_{[1,1,1]} = \frac{2}{z_{[1,1,1]}} : \hat{\mathcal{C}}_{1}^{3} : = \frac{1}{3} \hat{\mathcal{C}}_{1}^{3} - \hat{\mathcal{C}}_{1}^{2} + \frac{2}{3} \hat{\mathcal{C}}_{1} \\
    \ldots
    \end{gathered}
\end{equation}
Using \eqref{action_D} and \eqref{action_C1} one may easily reproduce the $\hat{\mathcal{W}}$-operators from \cite{MMNspin,MMZh}:
\begin{equation}
\begin{gathered}
    \hat{\mathcal{W}}_{[1]} = 2 \sum_{n \in \mathbb{Z}_{odd}^{+}} n P_{n}  \frac{\partial}{\partial P_{n}} = 2 \sum_{a=0}^{\infty} (2a+1) P_{2a+1} \frac{\partial}{\partial P_{2a+1}} \\
    \hat{\mathcal{W}}_{[1,1]} = \sum_{n,m \in \mathbb{Z}_{odd}^{+}} n m P_{n} P_{m} \frac{\partial^{2}}{\partial P_{n} \partial P_{m}} + \sum_{n \in \mathbb{Z}_{odd}^{+}} n(n - 1) P_{n} \frac{\partial}{\partial P_{n}} \\
    \ldots
\end{gathered}
\end{equation}
For other diagrams $\Delta$, one needs to apply also the second formula in (\ref{action_D}), the calculations
are equally straightforward, but lengthy, and we omit them from the present short text.
The first non-trivial  is labeled by partition $\Delta=[3]$:
\begin{equation}
    \hat{\mathcal{W}}_{[3]} = \frac{2}{z_{[3]}} : \hat{\mathcal{C}}_{3} : = \frac{2}{3} \hat{\mathcal{C}}_{3} - \frac{2}{3} \hat{\mathcal{C}}_{1}^{2} + \frac{1}{3} \hat{\mathcal{C}}_{1}
\end{equation}
and the answer coincides with that obtained in \cite{MMNspin,MMZh}.

The main lesson from this section is that,
in these time-variables, one can forget about the underlying superalgebras and fermionic variables:
we have just time-dependent eigenfunctions of time-dependent differential operators.
Still, one can not understand the structure of these quantities and explicit formulas for them
without understanding their relation to Sergeev characters.

\section{Algebraic description of results}

Let us put our results in more algebraic terms. One can start from the group algebra $A_N$ of the symmetric group $S_N$, which plays an essential role in describing the Schur-Weyl duality,
i.e. representations both of the symmetric group and of the matrix group $GL(N)$ --
and consider its counterpart $A_\infty$ for the infinite symmetric group. It is an algebra of the conjugated classes of finite permutations of an infinite set \cite{IK}, explicitly realized by $\phi_R(\Delta)$, see (\ref{IK}), which can be described \cite{IK,MMNspin} by the shifted Schur functions \cite{Oks}. In \cite{MMN2}, see sec.2, an exact representation of the algebra $A_\infty$ was found in the algebra of differential operators of infinitely many variables, obtained in the limit of $N\to\infty$ of the operators in the regular representation of the $N\times N$ matrix group.

Similarly, in \cite{MMNspin} we discussed the group algebra $As_\infty$ of the Sergeev group, explicitly realized by $\Phi_R(\Delta)$, see (\ref{MMNspin}), which can be described \cite{MMNspin} by the shifted Schur Q-functions \cite{I,Orlov,MMNspin,MMZh}. In this letter, we constructed an exact representation of the algebra $As_\infty$ in the algebra of differential operators of infinitely many variables obtained in the limit of $N\to\infty$ of the operators in the regular representation of the $2N\times 2N$ matrix supergroup. This can be considered as the main result of the paper.

\section{Conclusion}

The purpose of this letter was to extend the basics of cut-and-join calculus \cite{MMN1,MMN2}
to the Schur Q-functions.
Detailed explanations and immediate
applications to various matrix models will be provided elsewhere.
As clear from sections 5-6,
the actual calculations with Q-functions and Sergeev characters
are quite interesting but somewhat technical and sophisticated.
Therefore, we have found it reasonable to extract the concise and clear outcome
(\ref{PXPsi}), (\ref{WviaXPsi}), and to present it in a short focused text like this.

In this paper, we showed that the spin generalized cut-and-join $\hat{\cal W}$-operators
recently found in terms of time-derivatives \cite{MMNspin,MMZh},
can be realized as supermatrix differential operators (\ref{WviaXPsi})
after a proper Miwa parametrization of time-variables (\ref{PXPsi}), which in this case involves a fermionic matrix $\Psi$ in addition to $X$.
Note that in this realization times are {\it not} just traces of a supermatrix, but involve
{\it two} supermatrices differing by a sign of $\Psi$-component.

Our results have one more interesting implication.
The spin Hurwitz numbers and the Schur Q-functions are related \cite{MMNspin,MMNO} to the BKP integrable hierarchy \cite{OrlovBKP} in the same way as the ordinary Hurwitz numbers and the Schur functions are related \cite{GKM2,Ok,OS,Lando,MMN1,AMMN} to the KP hierarchy \cite{DJKM}.
What we demonstrated now is that they are also related to  supermatrices, thus, probably, to some super-hierarchies. This issue is also left for a future research.

\section*{Acknowledgements}

We are indebted to \boxed{\rm S.Natanzon}
and A.Orlov, who were the driving forces
in the search of the algebraic description
of Schur Q-functions and of the corresponding operators $\hat{\cal W}$.

This work was supported by the Russian Science Foundation (Grant No.20-71-10073)

\end{document}